\documentclass[prl,twocolumn,superscriptaddress,showpacs]{revtex4-1}
\usepackage{amsmath}
\usepackage{amssymb}
\usepackage{upgreek}
\usepackage{graphicx}
\usepackage{mathrsfs}
\usepackage[utf8]{inputenc}

\begin{document}

\title{Dynamic reorganization of vortex matter into partially disordered lattices}

\author{M. Marziali Bermúdez}
\affiliation{Departamento de Física, Facultad de Ciencias Exactas y Naturales, Universidad de Buenos Aires, Argentina.}
\affiliation{Instituto de Física de Buenos Aires, Consejo Nacional de Investigaciones Científicas y T{\'e}cnicas, Argentina.}

\author{M. R. Eskildsen}
\affiliation{Deptartment of Physics, University of Notre Dame, Notre Dame, IN 46556, USA.}

\author{M. Bartkowiak}
\affiliation{Laboratory for Developments and Methods, Paul Scherrer Institut, CH-5232 Villigen, Switzerland.}

\author{G. Nagy}
\affiliation{Laboratory for Neutron Scattering and Imaging, Paul Scherrer Institut, CH-5232 Villigen, Switzerland.}

\author{V. Bekeris}
\affiliation{Departamento de Física, Facultad de Ciencias Exactas y Naturales, Universidad de Buenos Aires, Argentina.}
\affiliation{Instituto de Física de Buenos Aires, Consejo Nacional de Investigaciones Científicas y T{\'e}cnicas, Argentina.}

\author{G. Pasquini}
\email{pasquini@df.uba.ar}
\affiliation{Departamento de Física, Facultad de Ciencias Exactas y Naturales, Universidad de Buenos Aires, Argentina.}
\affiliation{Instituto de Física de Buenos Aires, Consejo Nacional de Investigaciones Científicas y T{\'e}cnicas, Argentina.}

\date{\today }

\pacs{74.25.Uv, 61.05.fg, 64.60.Cn}

\begin{abstract}
We report structural evidence of dynamic reorganization in vortex matter in clean $\mathrm{NbSe_2}$ by joint small angle neutron scattering and ac-susceptibility measurements. The application of oscillatory forces in a transitional region near the order-disorder transition results in robust bulk vortex lattice configurations with an \emph{intermediate} degree of disorder. These dynamically-originated configurations correlate with \emph{intermediate} pinning responses previously observed, resolving a long standing debate regarding the origin of such responses.

\end{abstract}

\maketitle

In a wide variety of complex systems, competing interactions promote an order-disorder transition (ODT). Ordered phases are characterized by spatial correlations decaying weakly over distances larger than the relevant system scale, whereas disordered configurations are characterized by correlation lengths $\zeta$ of the order of the mean inter-particle distance $a_0$. Configurations with intermediate degrees of disorder, strong enough to affect the system response, but still with $\zeta \gg a_0$, have received attention recently\cite{packing}.
Vortex matter in superconductors provides an ideal model system for the experimental study of the topic \cite{Suderow2014}: vortex-vortex interactions favoring an ordered vortex lattice (VL) compete with both thermal fluctuations and pinning interactions that tend to disorder the system. 

In very low-pinning superconductors, such as clean $\mathrm{NbSe_2}$ single crystals, most of the vortex field-tem\-per\-a\-ture phase diagram is properly described by an ordered dislocation-free Bragg Glass (BG) phase \cite{BG1, BG2, Andrei2007}, which undergoes an ODT near the normal-superconductor transition \cite{ODT1, ODT2, ODTN1}. In practice, when a superconductor is cooled from the normal state in an external magnetic field (field-cooled/FC), energy barriers may trap the VL in highly disordered metastable configurations \cite{ODTN1}. Even so, high transport current densities \cite{HE1, Paltiel2000} or large oscillatory ``shaking'' magnetic fields \cite{Pasquini2008, Daroca2011} may anneal the VL into the ordered low-temperature BG.
Once in the BG, as temperature or magnetic field is increased the VL softens and accommodates to the random pinning potential more efficiently. Eventually, vortex entanglement and the proliferation of VL dislocations increase the effective pinning, producing a sudden rise of the critical current $J_{c}$ known as the Peak Effect (PE) \cite{ODT2, ODTN1}, which is the fingerprint of the ODT in vortex matter. 
While both the high-pinning disordered phase and the low-pinning ordered phase, above and below the ODT, are widely accepted in the literature, intermediate responses have been ascribed to surface contamination and partial reordering by the probing transport current \cite{HE1, Paltiel2000, SN, SN2}.
However, in the narrow transitional region adjacent to the PE, for which a multidomain phase has been theoretically proposed \cite{Menon2012}, non-invasive techniques have shown that pinning can be \emph{partially decreased} or even \emph{increased} by applying dc currents \cite{Xiao2000, Xiao2004} or ac magnetic fields \cite{Pasquini2008}. The resulting intermediate responses are highly reproducible and independent of the previous history. Are these in-between responses originated from bulk VL configurations with an intermediate degree of disorder?

To address this question, we performed an experiment in a clean $\mathrm{NbSe_2}$ single crystal, combining small angle neutron scattering (SANS) with \emph{in-situ} linear ac susceptibility measurements.
In SANS imaging of a perfect triangular VL, first order Bragg peaks (BP) are expected to appear as six symmetric sharp spots, matching the reciprocal lattice vectors.
This still holds for an ideal BG, where elastic disorder leads to slow decay of the correlations \cite{BG2}, although recent simulations \cite{Forgan2008} have shown that elastic BG prefracturing would result in a slight broadening of the BPs.
Conversely, when disorder is dominated by dislocations, spatial correlations decay exponentially beyond a characteristic length $\zeta$ associated with the scale at which dislocations appear \cite{Giamarchi1995}.
In such cases BPs have a finite intrinsic width proportional to $\zeta^{-1}$. Experimentally observed BPs are further smeared by a convolution kernel due to finite resolution \cite{ReviewMorten}.
Hence, the degree of disorder can be determined from the spread of the scattered intensity, as long as $\zeta \lesssim L_{res}$, a resolution-dependent bound.
In a 3D vortex system, the anisotropy of the elastic constants gives rise to very different characteristic lengths along the direction of the flux lines ($\zeta_L$) and in the transverse plane ($\zeta_{\perp}$).
Because of the poorer in-plane resolution, in this experiment only the longitudinal $\zeta_L$ was determined by measuring the integrated intensity on the plane of a single BP as the sample was rocked through the Bragg condition (rocking curve/RC).

\begin{figure}[tb]
\centering
\includegraphics[width=0.99\linewidth]{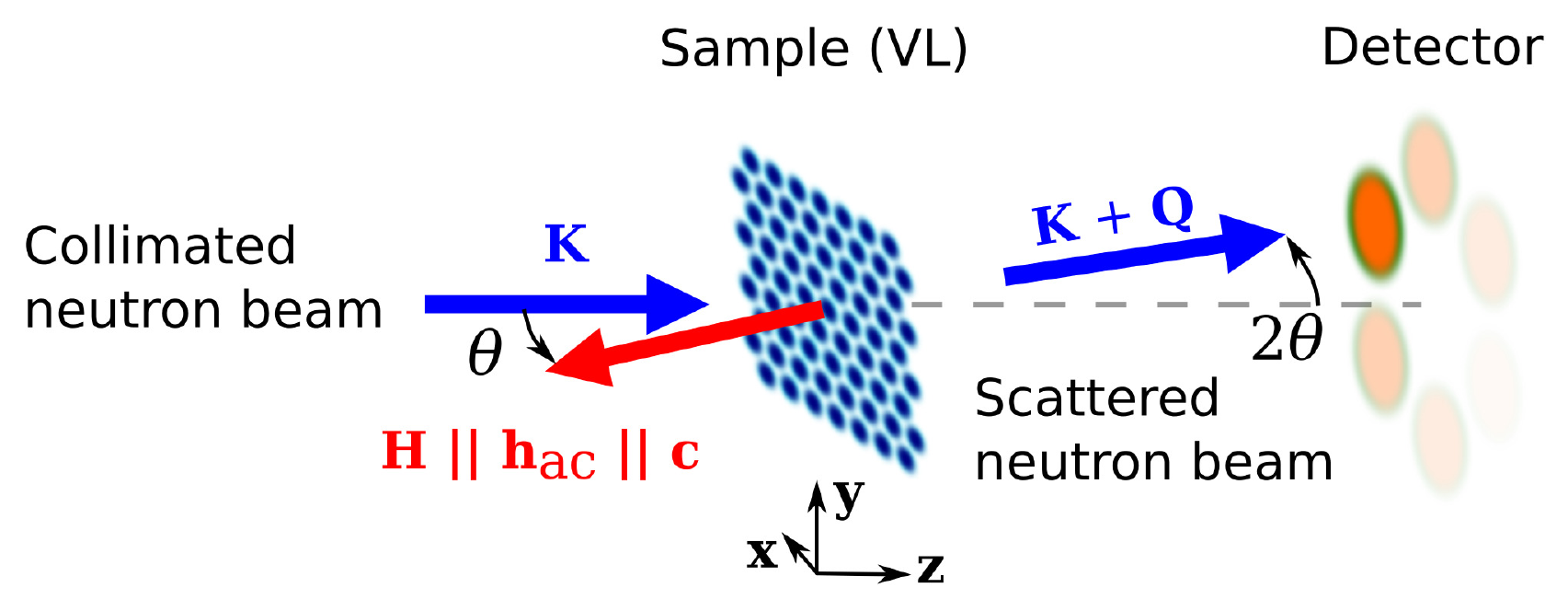}
\caption{
Scattering geometry: Both the dc field $\mathbf{H}$ and the ac field $\mathbf{h}_{ac}$ are aligned with the $\mathbf{c}$ axis of the single crystal, in the $\mathbf{xz}$ plane. 
As the whole system is \emph{rocked} around the $\mathbf{y}$ axis, the distribution of the scattered intensity $I$ is measured by the position sensitive detector, which spans the $\mathbf{xy}$ plane, for each angle $\theta$ between the field direction and incoming neutron beam wave vector $\mathbf{K}$. The maximum intensity is expected at $\theta_0 = Q_0/K$, for the corresponding reciprocal lattice vector $\mathbf{Q}_0$.
}
\label{fig1}
\end{figure}

The experiment was conducted at the SANS-II beam line of the Paul Scherrer Institute's SINQ facility, using the MA11 cryomagnet.
The geometry of the experiment is sketched in Figure \ref{fig1}.
All SANS measurements were performed at $T_0 = 1.9~\mathrm{K}$. Neutron wavelength was chosen to be $\lambda_{N} \approx 9~\mathring{\mathrm{A}}$ ($\pm 10\%$) and the applied dc field was $H=5~\mathrm{kOe}$. The incident beam was collimated, resulting in a beam divergence $\sim \pm 0.05~\mathrm{deg}$. The angular resolution along the rocking angle was estimated to be $w_{res}=(0.107 \pm 0.004)$~deg. A statistical analysis showed that curves arising from $\zeta_{L} < L_{res} \sim 40~\upmu\mathrm{m}$ were distinguishable from the resolution kernel. RCs wider than our resolution were fitted by the convolution of the Gaussian resolution with a Lorentzian distribution corresponding to an intrinsic $\Delta q_{L} = 2/\zeta _{L}$ (see Appendix).
A large ($5.6 \times 4.7 \times 0.2\,\mathrm{mm}^{3}$) clean $\mathrm{NbSe_2}$ single crystal was used for the experiment. The sample is part of a batch grown in Bell Labs as described in Ref. \cite{BellLabs}, with $T_{c}=(7.115 \pm 0.029)~\mathrm{K}$ ($10-90\%$ of the linear ac susceptibility transition at $H=0$). The phase diagram for various crystals has been built using a 7-T MPMS~XL (Quantum Design) and compared with those published in the literature \cite{Xiao2004}, showing non significant differences (see Appendix).

Since the goal was to explore the connection between structure and dynamics, a non-invasive measurement of pinning was required. This was achieved by installing a sample holder with mutual inductance coils inside the SANS cryomagnet. By adding a small ac magnetic field $h_{ac}=2.5~\mathrm{mOe}$ ($< 10^{-6}~H, f=65~\mathrm{kHz}$), vortices were forced to perform small oscillations inside their pinning wells, without modifying their spatial configuration. These oscillations propagate through the sample due to the vortex-vortex repulsion, with a characteristic penetration depth $\lambda _{ac}$ that is related, in a mean field approximation \cite{Colo2014}, to the effective pinning potential \cite{Campbell}. The linearity of the response, as well as the low dissipation level characteristic of the linear Campbell response \cite{Campbell} had been previously verified. Then $\lambda_{ac}$ and thus the effective pinning can be assessed through the in-phase component of the linear ac susceptibility $\chi'$ by means of a mutual inductance technique. Lower values of $\chi'$ are associated with stronger pinning, saturating at perfect ac shielding (normalized to $\chi'= -1/4\pi$), whereas $\chi'$ vanishes under complete ac penetration when $\lambda_{ac} \gtrsim$~sample size. 
To probe oscillatory dynamic effects, a \emph{shaking} procedure was applied at certain temperatures: Here $\chi'$ measurements were paused and a larger sinusoidal field with amplitude $h_{sh}=7$~Oe, and $f_{sh}=1$~kHz was applied for 1000 cycles before returning to the smaller $h_{ac}$ to resume $\chi'$ measurements. The shaking field amplitude  was sufficient to induce vortex displacements larger than $a_{0}$ throughout the sample. The complete penetration of the shaking field into the sample had been previously confirmed by measuring the non-linear ac response and checking consistency with the irreversible dc magnetization determined from $M(H)$ loops  (see Appendix). 

\begin{figure}[b]
\centering
\includegraphics[width=0.99\linewidth]{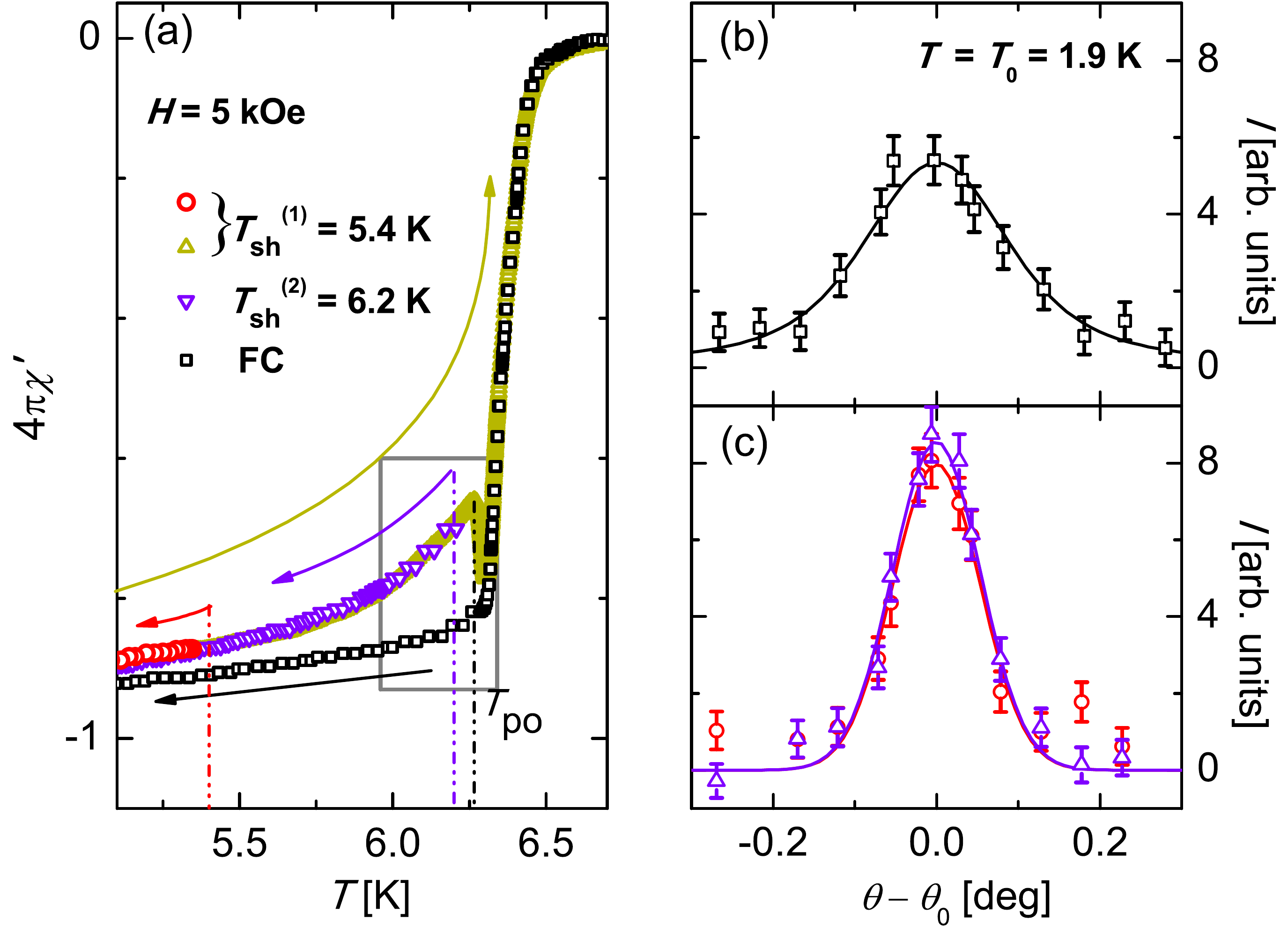}
\caption{
(a) Linear ac susceptibility $\chi'(T)$ measured down to $T_0 = 1.9~\mathrm{K}$ during a FC cooling procedure (black squares) and after shaking the VL at two different temperatures $T_{sh}^{(1,2)}$ (red circles and violet triangles) below the onset of the PE ($T_{po}$). Shaking temperatures are indicated by vertical lines and arrows indicate temperature evolution. Reversibility is observed in the warming from $T_0$ of the VL shaken at $T_{sh}^{(1)}$ (dark yellow triangles). The area enclosed by the rectangular box is magnified in Fig. \ref{fig3}(a).
(b) RC after FC; the black curve is the fit of a Lorentzian peak convoluted by the resolution kernel.
(c) RCs after shaking the VL at $T_{sh}^{(1)}$ (red circles) and $T_{sh}^{(2)}$ (violet triangles); the curves are Gaussian fits with FWHM equal to the instrumental resolution $w_{res}=0.107$ deg.
}
\label{fig2}
\end{figure}

Figure \ref{fig2}(a) shows a comparison between $\chi'(T)$ measured during a FC procedure and after shaking the VL at two different temperatures $T_{sh}^{(1,2)}$ below the onset of the PE at $T_{po} \simeq 6.27~\mathrm{K}$. There is an unambiguous rise in $\lambda_{ac}$ after shaking, representing a reduction of the effective pinning. The agreement between both cooling $\chi'(T)$ curves after shaking  is consistent with an equilibration procedure below $T_{po}$. Reversibility is observed in the warming curve, with $\chi'(T)$ increasing with $T$ up to $T_{po}$, after which it drops due to the spontaneous pinning enhancement at the PE. In contrast, the absence of PE in the linear response during the cooling \cite{Pasquini2008} is likely due to the VL remaining trapped in a metastable highly pinned state.
This explanation is consistent with the FC RC (Figure \ref{fig2}(b)), which is clearly wider than the resolution kernel, in agreement with other experiments \cite{Yaron1995}. Conversely, the two RCs recorded after shaking the VL (Figure \ref{fig2}(c)) show resolution-limited BPs ($\zeta_L > L_{res}$). The fact that BPs are well defined (unlike the annulus observed for amorphous VLs) indicates that orientational order is preserved in all the cases. The broadening of the FC configuration RC suggests a dislocated VL with broken positional order beyond a characteristic volume with longitudinal dimension $\zeta_{L} < L_{res}$. There is a clear correlation between the ordering of the VL and the decrease of the effective pinning.

\begin{figure}[tb]
\centering
\includegraphics[width=0.99\linewidth]{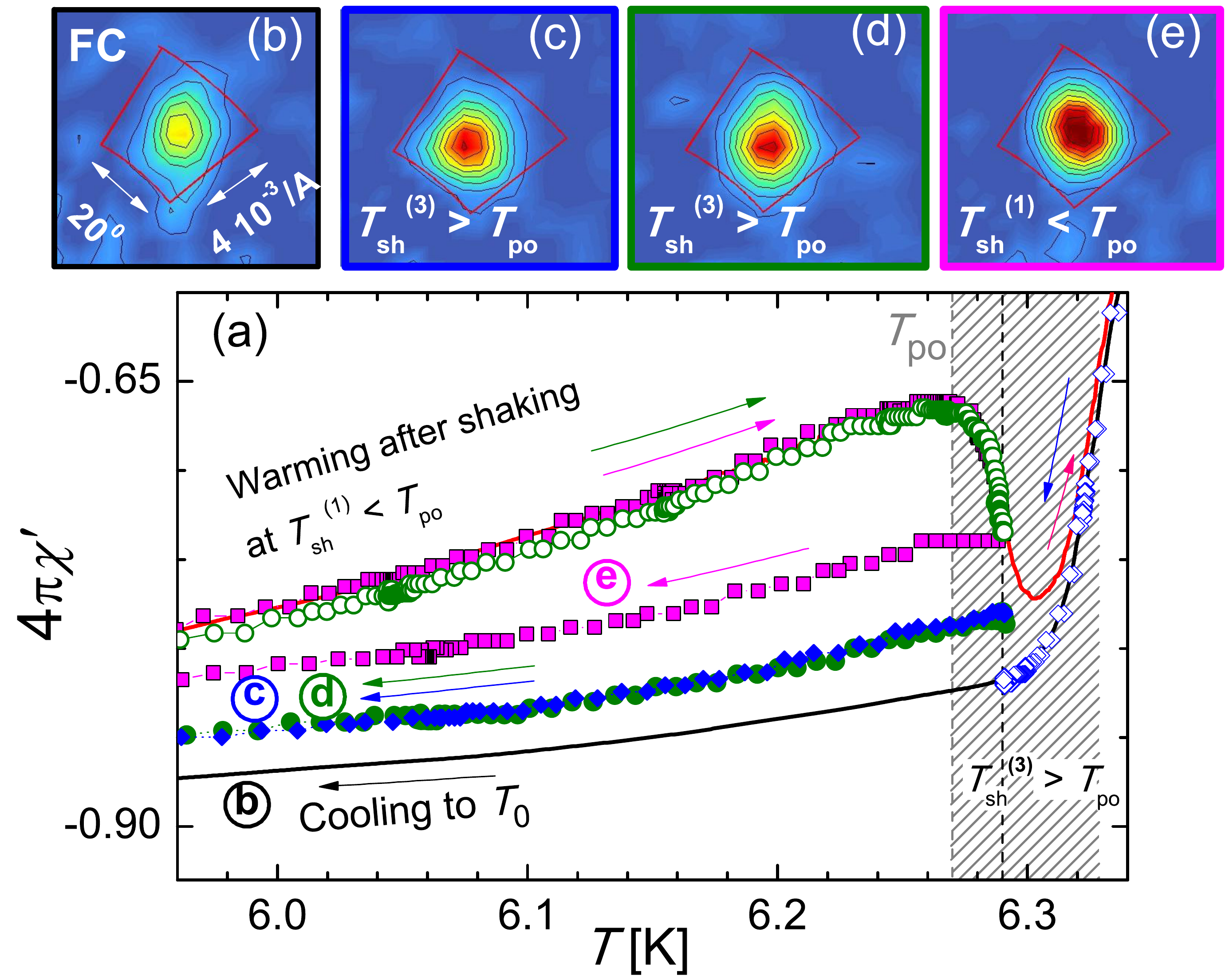}
\caption{
$\chi'(T)$ for different thermal and dynamic histories (labeled (b)-(e) in panel (a)) and the corresponding detector intensity distributions for one VL BP at the Bragg condition (panels (b)-(e)). Arrows in (a) indicate temperature evolution and the hatched area shows the transitional region. Data from Figure \ref{fig2}(a) (black and red lines) is included for comparison. 
The application of a shaking field at $T_{sh}^{(3)} \approxeq 6.29~\mathrm{K} > T_{po}$ results in a unique response (full blue and green symbols) independent of whether the  initial state was a disordered FC VL (open blue symbols) or an ordered VL (open green symbols). Both intensity distributions (c) and (d) indicate partial disorder as compared to the FC intensity distribution (b). Simply warming an ordered VL (magenta squares) up to $T \approxeq T_{sh}^{(3)}$ shows moderate pinning enhancement during cooling, but the corresponding BP (e) is still sharp and intense. The red boxes in (b)-(e) enclose the detector area included in the RCs in Figure \ref{fig4}.
} 
\label{fig3}
\end{figure}
\begin{figure}[!ht]
\centering
\includegraphics[width=0.9\linewidth]{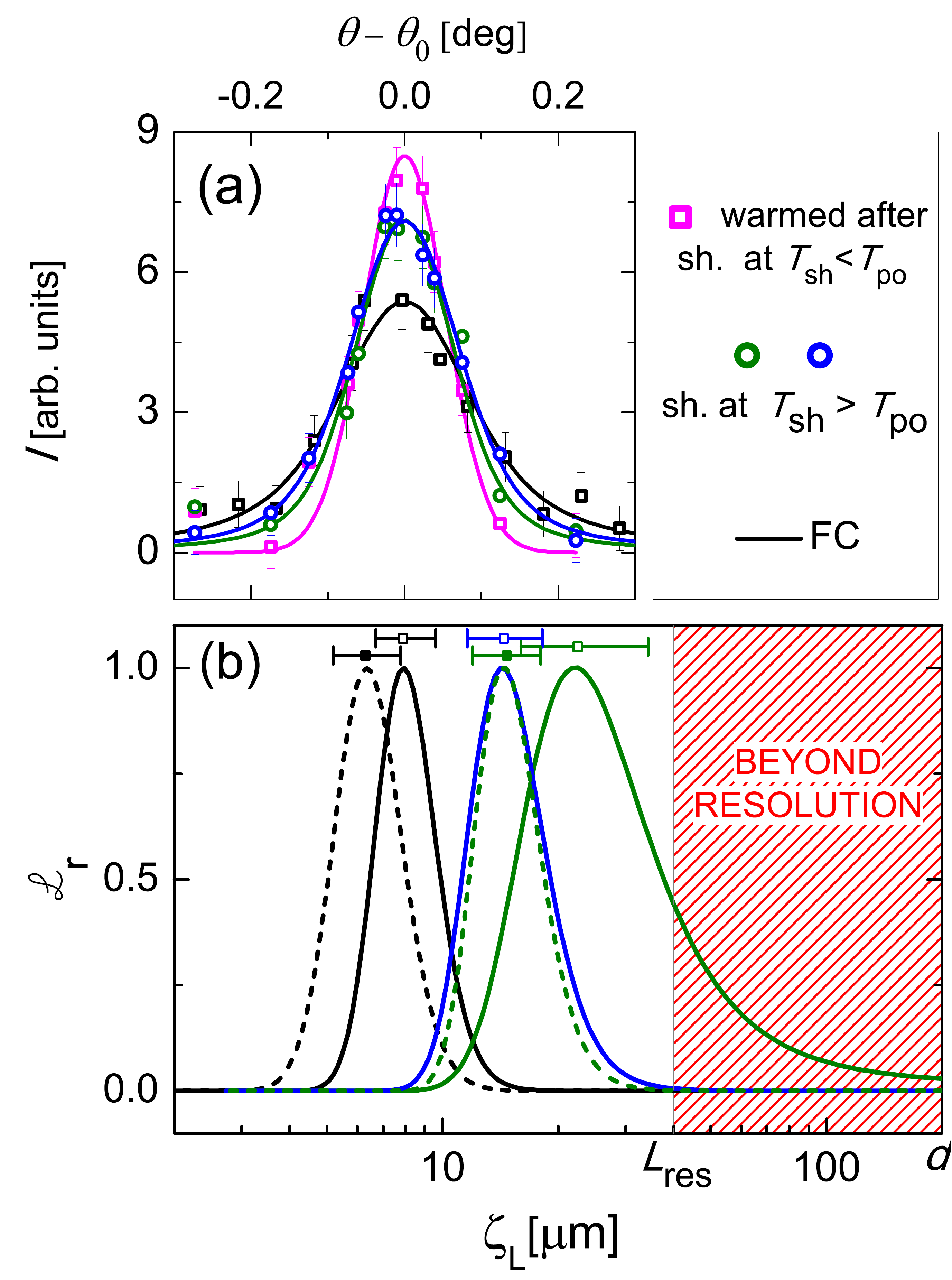}
\caption{
(a) Rocking curves measured after different field-temperature histories: FC (black, same as in Fig. \ref{fig2}(b)), after shaking at $T_{sh}^{(3)}$ a FC (blue) and an ordered (green) VL, and after warming an ordered VL to $T = T_{sh}^{(3)}$ without shaking at this temperature (magenta). Curves are fits of Lorentzian peaks convoluted by the resolution.
(b) Calculated profile likelihood ratio $\mathscr{L}_r$ as a function of $\zeta_{L}$ up to the sample thickness $d$ for each experimental RC. Symbols above the curves indicate the maximum ($\mathscr{L}_r \equiv 1$) and error bars, the $68\%$ confidence interval. For the ordered VL, $\zeta_{L} > L_{res}\sim 40~\upmu\mathrm{m} \sim 570~a_{0}$ and is beyond our resolution. The profile likelihood of replicate RCs of the FC (black dashed line) and the one shaken at $T_{sh}^{(3)} \pm 3~\mathrm{mK}$ (green dashed line) are reproducible within the error bars.
}
\label{fig4}
\end{figure}

While shaking the VL at any $T_{sh} < T_{po}$ leads to an ordered configuration, the consequences are strikingly different when the shaking field is applied in the narrow transitional region above $T_{po}$. Figure \ref{fig3}(a) shows the evolution of $\chi'(T)$ through a series of processes near the transitional region (hatched area) and Figure \ref{fig4}(a) displays the corresponding RCs.
A strongly pinned FC VL was cooled to $T_{sh}^{(3)}=6.29\,\mathrm{K} > T_{po}$ and shaken, partially reducing the effective pinning. The VL was cooled to $T_0$ and a narrow RC, still wider than $w_{res}$ was measured, indicating a bulk \emph{partial ordering}. In contrast, applying the same shaking burst at the same $T_{sh}^{(3)}$ ($\pm 3~\mathrm{mK}$) to an initially ordered VL had the opposite effect: The shaking field partially increased the effective pinning to a similar level and \emph{disordered} the VL. The RC was broader than $w_{res}$, but narrower than for the maximally disordered FC lattice. Because the response after shaking is strongly dependent on the temperature when $T_{sh} > T_{po}$, the process was carefully repeated within $\Delta T_{sh} \sim 3~\mathrm{mK}$, obtaining the same qualitative results. 
To rule out that the \textit{partial disordering} was of thermal origin, a warming-cooling cycle up to $T_{sh}^{(3)}$ but without applying the shaking burst was carried out for a replicated ordered VL. Although the cooling $\chi'(T)$ branch indicates higher pinning than the warming branch, it is lower than for VLs shaken above $T_{po}$. Moreover, if there was any associated disorder, it was minor and below our SANS resolution. Therefore, the ``intermediate'' degree of disorder found before is only attributable to the shaking field applied at $T_{sh}^{(3)}$.

Although the above results are qualitatively evident by simply comparing the corresponding intensity distribution on the position sensitive detector at $\theta =\theta_{0}$ (Figure \ref{fig3}(b)-(e)), we evaluated the statistical significance of the observed differences (see Appendix). Figure \ref{fig4}(b) shows the profile likelihood ratio $\mathscr{L}_r(\zeta_{L})$ resulting from the fit of each experimental RC to one expected from a configuration with an intrinsic width given by $\zeta_{L}$ convoluted by the experimental resolution.
Correlation lengths obtained after shaking the VL at $T_{sh}^{(3)}$ were $L_{res} > \zeta_{L} > \zeta_{L}(\mathrm{FC})$ at the $75\%$ confidence level. Thus, we state that shaking the VL in the transitional region results in bulk intrinsic configurations with intermediate degree of disorder.

The total integrated intensity $I_0$ (the area under the RC) for the different configurations should remain roughly constant. Even so, the long-tail decay of the structure factor in the BG phase may lead to an underestimation of $I_0$ \cite{BG2}. 
Indeed, for all FC configurations and for the configurations obtained after shaking the VL in the transitional region we observe a similar $I_0$. However, for the lattice shaken in the BG the integrated intensity is slightly, but statistically significant, smaller (see Appendix).
These features support a qualitative difference in the positional order decay in the elastic BG configurations and in those obtained after shaking the VL in the transitional region, where disorder is probably mediated by dislocations.

In the elastic BG, the in-plane correlation length $\zeta_{\perp}$ can be derived from $\zeta_L$ through the tilt and shear elastic constants $c_{44}$ and $c_{66}$ \cite{BG1,BG2,lambdaT,Brandt1995}. From $L_{res}$, a lower limit of $35~a_0$ can be estimated  for the in plane size of the dislocation-free regions corresponding to the ordered configurations, which is consistent with results obtained in other works in the BG phase \cite{BG2,Yaron1995,Ganguli2014}. How $\zeta_{\perp}$ relates to $\zeta_L$ is not as straightforward for metastable VL configurations with dislocations. In theoretical works \cite{Mullock, TheoWorks}, $\zeta_L$ and $\zeta_{\perp}$ are obtained for equilibrium VL configurations from the densities of screw and edge dislocations that minimize the free energy; far enough from the amorphous limit (i.e., $\zeta_{\perp} \sim a_0$) dislocation densities adapt in such a way as to satisfy the elastic ratio. Moreover, numerical simulations \cite{Kolton2000} and experimental results \cite{Yaron1994} show that $\zeta_L$ monotonically increases with the in-plane ordering from the plastic to the elastic regime. In our case, the partially disordered VLs are metastable configurations (at $T_0$) which originated from a dynamic reorganization in the vicinity of the ODT.  In this region the elastic constants are dispersive and may also be affected by fluctuations. Therefore, a precise quantitative estimation of $\zeta_{\perp}$ from $\zeta_L$ is not available, but a monotonic correlation is still expected (see Appendix for a quantitative estimate). Such a correlation would be compatible with an intermediate density of dislocations in partially disordered VLs.

Our results show a clear connection between the linear ac response, related to the effective pinning, and the bulk spatial correlation of the VL. Configurations having correlation lengths much larger than the lattice parameter, but small enough to affect the global response, were observed after shaking the VL in the proposed transitional region. These observations support the existence of such region and represent the first evidence of a dynamic reordering, driven by oscillatory forces, which results in robust bulk configurations with intermediate degrees of disorder. A similar behavior was obtained in numerical simulations \cite{Daroca2011}, where configurations with intermediate dislocation densities are accessible from stationary fluctuating dynamic states, as those proposed for colloids \cite{Pine2008, Reichhardt2009}. The interplay between these spatial configurations and their associated non-linear dynamics remains a fertile ground for further research.

\section*{Acknowledgements} 
This work is based on experiments performed at the Swiss spallation neutron source SINQ, Paul Scherrer Institute, Villigen, Switzerland, and was partially supported by Fundación J. B. Sauberán and ANPCyT under Grant PICT No. 753. MRE was supported by the U. S. Department of Energy, Basic Energy Sciences under Award No. DE-FG02-10ER46783.  We are grateful to G. Nieva for providing the $\mathrm{NbSe_2}$ sample, M. Kenzelmann for financial aid and E. R. De Waard for assistance with the SANS experiment.

\section*{Appendix}

\section{SANS data analysis}

\subsection{Resolution model}

We modeled rocking curves (RCs) as the convolution of an \emph{intrinsic} intensity distribution with a smearing \emph{resolution function}. The former corresponds to the actual $q$-dependence of the structure factor, while the latter accounts for imperfect collimation and wavelength spread. For simplicity, the resolution function was assumed to be a Gaussian kernel with standard deviation $w_{res}/2$. Since rocking curves are already integrated within the detector plane, only the longitudinal dependence of the structure factor is relevant.

When spatial correlations of the VL decay strongly with certain characteristic length scale, the envelope of the correlation function can be modeled with an exponential function. So, the structure factor $S(q_{L})$ (the Fourier transform of the correlation function) must be Lorentzian-shaped, i.e. 
\begin{equation}
S(q_{L})\propto \frac{1}{1+(\zeta_{L}q_{L})^{2}},
\end{equation}
where $\zeta_{L}$ is the corresponding correlation length. Therefore, we took the intrinsic distribution to be the Cauchy-Lorentz distribution centred at the Bragg angle $\theta_{0}$, multiplied by a proportionality constant $I_{0}$. Then, its half width at half maximum (HWHM) $\gamma$ is directly related to $\zeta_{L}$ by 
\begin{equation}
\gamma =\frac{1}{Q_{0}\zeta_{L}},
\label{corrl}
\end{equation}
where $Q_{0}$ is the magnitude of one reciprocal vector of the VL.
The resulting intensity profile is 
\begin{equation}
I(\theta ) = \int_{-\infty }^{\infty }\frac{I_{0}}{\pi }\frac{\gamma }{\gamma^{2}+(\theta ^{\prime }-\theta_{0})^{2}}\frac{2\, e^{-2\left( \frac{\theta ^{\prime }-\theta }{w_{res}}\right) ^{2}}}{\sqrt{2\pi }w_{res}} \mathrm{d}\theta ^{\prime }.
\label{intdist}
\end{equation}
In the limiting cases of long correlation lengths $Q_{0}\zeta_{L}\gg 1/w_{res} $, (i.e. when $\zeta_{L}$ is beyond our resolution) the intrinsic distribution can be replaced by a Dirac delta, so 
\begin{equation}
I(\theta )\approxeq \frac{2I_{0}}{\sqrt{2\pi }w_{res}}\exp \left[- 2\left( \frac{\theta -\theta_{0}}{w_{res}}\right) ^{2}\right] .  \label{longcorr}
\end{equation}
and the experimental RCs can be just modeled by the Gaussian resolution distribution.

\subsection{Data reduction}

Neutron count data were preprocessed by the software GRASP \cite{grasp}. Detector counts were normalized by monitor counts. Then, after background subtraction, intensity integrated over a fixed sector of the detector was calculated for each image, and resulting rocking curves $(\theta_j, I_j)$, $j = 1 \dots N$, were exported, together with their error estimates $\Delta I_j$.

\subsection{Data analysis}

Statistical analysis was carried out using the software Octave \cite{octave}. It consisted of three stages:
\begin{enumerate}
\item preliminary analysis,
\item resolution estimation,
\item intrinsic widths estimation.
\end{enumerate}

\subsubsection{Preliminary analysis}

Non linear optimization often requires set of initial parameters close enough to the optimal in order to work properly. The aim of this stage was to obtain suitable estimates of the parameters for each rocking curve individually, to be used in later stages. We did this by fitting our convolution model individually to each rocking curve.

Because of the convolution, the parameters $w_{res}$ and $\gamma$ are expected to be highly correlated, making it difficult to fit both simultaneously. To overcome this issue, a rough prior theoretical estimate of $w_{res}$, based on the experimental geometry \cite{morten}, was calculated. This resulted in $\tilde{w}_{res} = (0.104 \pm 0.003)\,\mathrm{deg}$. A modified likelihood function $\tilde{\mathscr{L}}$ was then constructed and parameters were obtained from minimizing 
\begin{equation}
\begin{split}
-2 \log \tilde{\mathscr{L}}(I_0,\theta_0,\gamma,w_{res}) &= \\ \left(\frac{w_{res} - \tilde{w}_{res}}{\Delta \tilde{w}_{res}}\right)^2 &+ \sum_{j=1}^N \left(\frac{I_j - I(\theta_j)}{\Delta I_j}\right)^2.
\end{split}
\end{equation}

\subsubsection{Resolution estimation}

The preliminary analysis showed that configurations shaken at low temperatures (examples are shown with red and violet symbols in Figure \ref{fig2}(c)) were resolution limited ($\gamma \ll w_{res}$). Therefore, we used the long-correlation-length approximation (Eq. \ref{longcorr}) to model these rocking curves. In turn, it enabled us to make an independent estimation of the resolution based solely on neutron data.

The resolution profile likelihood $\mathscr{L}_w(w_{res})$ was constructed by maximizing the global likelihood function (i. e., including all rocking curves), constrained to $w_{res}$ taking a series of fixed values. We obtained $\hat{w}_{res}=(0.107\pm 0.004)\,\mathrm{deg}$, which is consistent with our prior estimate. This also revealed that the dependence of the estimated $\gamma $ on the resolution width was common to all the curves. Examples of the resulting fitting curves are shown in full red and violet lines in Figure \ref{fig2}(c).

Next, we determined the minimum intrinsic width $\gamma $ which could be eventually distinguishable from the zero-width approximation. This was done by generating randomized rocking curves of different widths with the convolution model, using the above estimation of $w_{res}$. Similar number of points and relative errors as in our experimental data were considered. After fitting Gaussian functions to generated data, a threshold of $>90\%$ out of 200 replicates resulting in non-overlapping error bars (one standard deviation) between $\hat{w}_{res}$ and the fitted width was applied. This resulted in $\gamma \gtrsim 0.014\,\mathrm{deg}$, or $\zeta_{L}\lesssim 40\,\mathrm{\upmu m}$ $\sim L_{res}$ (vertical dashed line in Figure \ref{fig4}(b)).

\subsubsection{Intrinsic correlation length estimation}

Lastly, the profile likelihood of $\gamma $ for each rocking curve was constructed by fitting only $I_{0}$ and $\theta_{0}$, while letting ${w}_{res}$ be the estimated above (fixed). This procedure reduced marginal errors in $\gamma $ by ignoring the covariance with ${w}_{res}$. Since a change in ${w}_{res}$ would only shift all estimated $\gamma $ together, this approach is better suited for comparing different configurations instead of finding the exact correlation length. The resulting likelihood ratios $\mathscr{L}_r(\zeta_{L})$ (curves in Figure \ref{fig4}(b)) where then obtained through the relationship between $\zeta_{L}$ and $\gamma $ (Eq. \ref{corrl}).

As long as the elastic approximation is valid, the in-plane characteristic lengths $\zeta _{\perp }$ in equilibrium configurations can be estimated from the elastic ratio \cite{BG1,BG2} 
\begin{equation}
\zeta _{\perp }\sim \zeta _{L}\sqrt{c_{66}/c_{44}},
\end{equation}
where $c_{66}$ and $c_{44}$ are the shear and tilt elastic VL moduli respectively. In the local approximation \cite{Brandt1995}, 
\begin{equation}
c_{44}/c_{66}\sim 16\pi \ (\lambda /a_{0})^{2},
\end{equation}
where $\lambda $ is the London penetration depth and $a_{0}$ is the mean intervortex distance. In our case \cite{lambdaT}, $\lambda (T=2$ K$)/a_{0}\sim 2$, so the corresponding in-plane characteristic lengths in the BG phase are expected to be $\zeta _{\perp BG}>L_{res}/14\sim 35~a_{0}$. \ On the other hand, the fact that all the BPs are resolution limited, indicates that all the measured VL configurations are not in the amorphous limit.

Far from the amorphous limit, the elastic approximation is still roughly valid for equilibrium configurations with dislocations \cite{Mullock,TheoWorks}. However, as mentioned in the main text, the partially disordered VLs are metastable configurations at $T_{0}=2K$, originated from a dynamic reorganization in the vicinity of the ODT, and therefore a quantitative calculation of $\zeta_{\perp }$ from $\zeta_{L}$ is not available. A rough estimation using the elastic limit at $T_{0}$ (where the local approximation for the elastic constant is valid) gives $\zeta _{\perp }(\mathrm{FC}) \sim 5~a_{0}$ and $\zeta_{\perp }\sim 12~a_{0}$ for configurations with intermediate disorder. A better estimate can be made in the framework of a defective VL \cite{Mullock,TheoWorks} at high temperature (where the dynamic reorganization occurs). In this model each correlated volume $\zeta_{L}\zeta _{\perp }^{2}$ is limited by dislocations spaced in $\zeta_{\perp}$ and $\zeta_{L}$ in the transverse and longitudinal directions respectively. To calculate $\zeta_{\perp }(\zeta_{L})$ we follow Eqs. (9) and (10) of Ref. \cite{Mullock} (where the notation is $D_{c} \equiv \zeta_{\perp }$, $L_{c} \equiv \zeta _{L}$) and a non local approximation for the tilt modulus \cite{Brandt1995} $c_{44}(\zeta_{\perp}) \sim c_{44}(0)/(\lambda ^{2}k^{2}+1) \sim c_{44}(0)/(2\pi \lambda / \zeta_{\perp })^{2}$ where the penetration depth $\lambda \gg \zeta_{\perp}/2\pi $. The resulting $\zeta_{\perp}(\zeta_{L})$ are $\lambda$-independent in this limit and the in-plane correlation lengths corresponding to the experimental $\zeta _{L}$ result $\zeta _{\perp }(\mathrm{FC})\sim 9~a_{0}$ and $\zeta _{\perp }\sim 18~a_{0}$ for intermediate disordered configurations, slightly larger to that obtained in the elastic limits. Although these values are not quantitatively exact, these estimates could be useful to design future SANS experiments.

\subsection{A comment on total integrated intensities}

Equation (\ref{intdist}) is properly normalized so that the parameter $I_0$ represents the total integrated intensity. The maximum-likelihood estimate for $I_0$ and the corresponding variance were obtained along with the other parameters discussed above for each RC. In order to compare integrated intensities corresponding to different configurations, the dispersion in the intensities fitted to RCs measured after different histories were statistically assessed by means of $\chi^2$ difference tests of nested models. The results of the tests are presented indicating the statistic $\chi^2$, the degrees of freedom $df$, and the corresponding $p$-value (i. e., the probability of getting a $\chi^2$ greater than the observed one solely due to randomness, under the assumption that a common mean is shared).

First we note that the set of all the values of $I_0$ had far more dispersion (relative to their errors) than expected for a single mean value model ($\chi^2= 42.48$, $df = 10$, $p < 0.001$). In order to assess whether this excessive dispersion was connected to the different histories, the dispersion within sets of $I_0$ corresponding to each process was evaluated. All intensities corresponding to FC configurations were indistinguishable ($\chi^2= 1.37$, $df = 2$, $p = 0.505$) and intensities corresponding to intermediate-order configurations were only marginally distinguishable ($\chi^2= 6.15$, $df = 2$, $p = 0.046$). Moreover, merging the sets of intensities for FC and intermediate-order configurations did not increase the dispersion enough to reject a single mean intensity for both ($\Delta \chi^2 <0.01$, $\Delta df = 1$, $p = 0.938$). On the contrary, BG configurations presented a higher degree of variability, beyond their estimated variance ($\chi^2= 14.78$, $df = 4$, $p = 0.005$) and the set of intensities for BG configurations could not be merged with FC and intermediate-order configurations without increasing the dispersion significantly ($\Delta \chi^2 =20.18$, $\Delta df = 1$, $p < 0.001$).

These results suggest that all RCs for FC and intermediate-order configurations share essentially the same value for the parameter $I_0$. Instead, fits to RCs for BG configurations yielded statistically distinguishable values for the total integrated intensity, which averaged about $19\%$ below the mean $I_0$ for FC and intermediate-order configurations. As mentioned in the main text, an underestimation of $I_0$ is likely to result from the long-tail decay of the structure factor in the BG phase, further supporting the observed structural differences.

\section{Vortex phase diagram}

\begin{figure}[b]
\centering
\includegraphics[width=\linewidth]{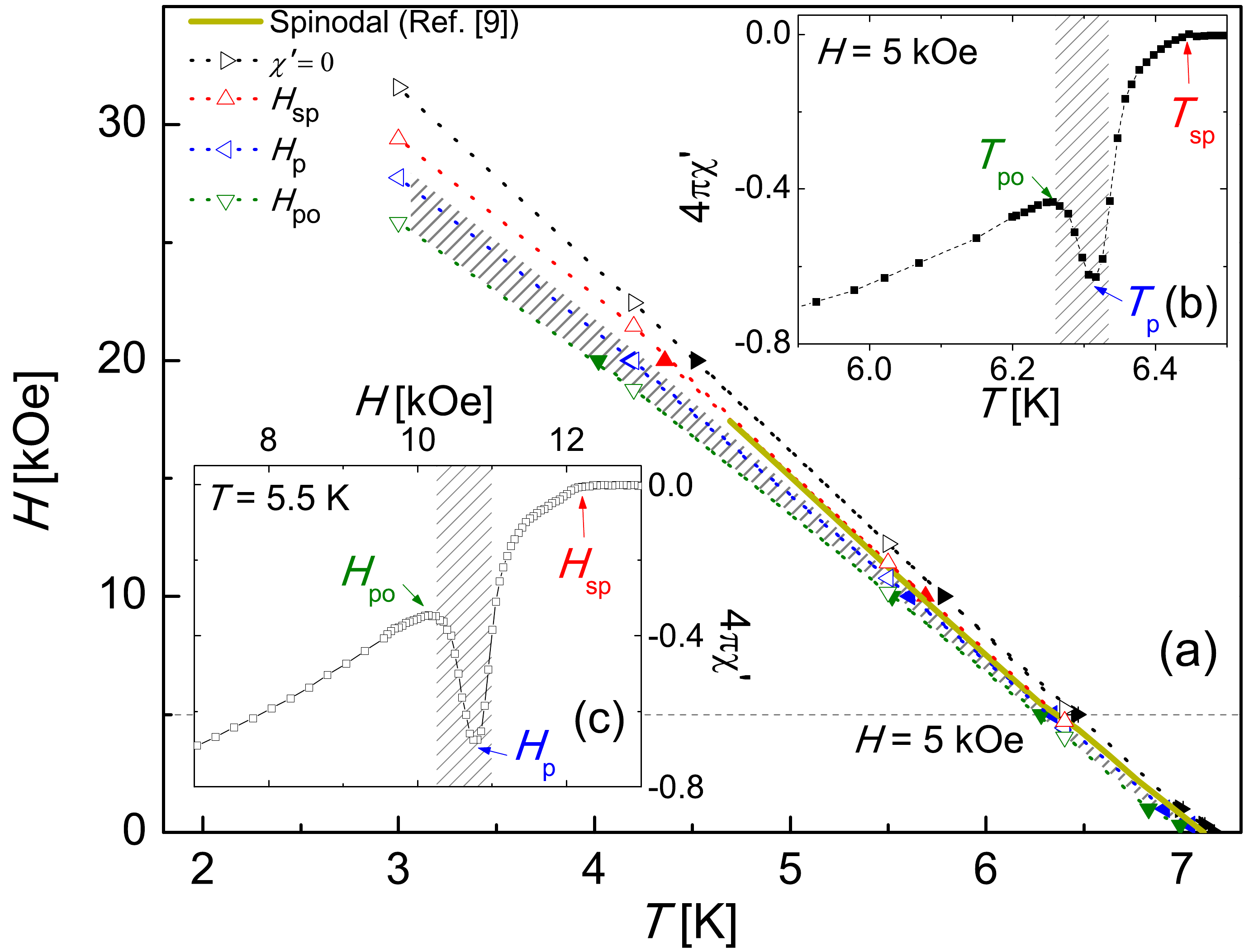}
\caption{(a) Vortex phase diagram obtained from $\chi'(T)$ (full symbols, example in panel (b)) and $\chi'(H)$ (open symbols, example in panel (c)). Below $H \sim H_{po}(T)$ a shaking field orders the VL. The hatched region indicates the transitional region, where we claim that the VL dynamically reorganizes in configurations with intermediate degree of disorder. Above $H \gtrsim H_{p}(T)$ there is a narrow region where disordered VLs are stable. No history effects are observed beyond $H_{sp}(T)$, which matches the spinodal line described in Ref. \cite{Xiao2004} (dark yellow line).}
\label{fig:s1}
\end{figure}

In this supplementary section, the vortex phase diagram corresponding to our samples is presented and some features are compared with those reported in the literature. The magnetic field and temperature ranges used in our experiments are also included. As the sample used in the SANS experiment (in the following, sample A) was too large to be measured in the 7-T MPMS XL Quantum Design or in our cryostats, another crystal from the same source (sample B, $V\sim 0.25\times 0.25\times 0.02$ cm$^{3}$) and a smaller sample obtained by cutting an edge of sample A (sample C, $V\sim 0.1\times 0.1\times 0.02$ cm$^{3}$) were characterized instead. Beyond the expected differences due to the larger area of sample B, results obtained from both samples are identical. 
In Figure \ref{fig:s1}, the phase diagram obtained from non-linear ac susceptibility curves $\chi'(H)$ (open symbols) and $\chi'(T)$ (full symbols) measured in sample B is shown in panel (a), including the transitional region (hatched area in Fig. \ref{fig:s1}). The dark yellow line is the spinodal, lying above the transitional region, that demarcates the limit of stability with respect to fluctuations toward the thermodynamically stable state, extracted from Ref. \cite{Xiao2004} and rescaled using the critical parameters of our sample $T_{c}=7.11$~K and $H_{c2}(0)=54.45$~kOe (estimated by extrapolating the line $\chi'(H,T)=0$ to $T=0$). This line matches the abrupt change in slope observed in both $\chi'(T)$ ($T_{sp}$ in panel (b)) and $\chi '(H)$ ($H_{sp}$ in panel (c)), above which no history effects are observed. The fact that our experiments are performed below the spinodal line indicates that fluctuations are not dominant in the underlying physics. However, due to the close vicinity, some small influence cannot be discarded. The dotted horizontal line indicates the magnetic field applied in our SANS experiment.

\section{Magnetic field penetration and vortex displacements}

The full bulk penetration of the shaking fields, with the consequent vortex displacements in distances larger than $a_{0}$, is essential to justify our claim of a bulk dynamic reordering. The modification of the bulk VL structure factor observed in the SANS experiment is the main direct evidence of  such a reordering. In this supplementary section, we present some numerical estimations, together with additional experimental results obtained by magnetization and non-linear ac susceptibility measurements, that provide a consistent scenario regarding this crucial issue.

\subsection{Estimation of vortex displacements}

A simple calculation provides a rough estimation of the mean vortex displacement associated with the shaking field penetration. Under full ac field penetration, the maximum variation of the VL parameter $a_{0}$ during a cycle of the shaking field is
$$\Delta a_{0}\sim (a_{0}/2)(h_{sh}/B)\sim (a_{0}/2)(h_{sh}/H).$$ 
In our SANS experiments, $h_{sh}\sim 7$~Oe and $H=5$~kOe, so $\Delta a_{0}\sim 0.05$~nm. Therefore, near the centre of the sample, vortices only need to move short distances $u \gtrsim \Delta a_{0}$ to accommodate to the change in the density.  However, the propagation of the density change throughout the sample implies that the mean displacement increases with the distance $r$ from the centre as $u(r)\sim r\ (\Delta a_{0}/a_{0})$ . Consequently, in our experiments we expect $u(r) > a_{0}$ for $r>100$~$\upmu$m, reaching $u(r)$ as large as several microns near the edge of the sample ($r\sim 2.5$ mm). In conclusion, during the shaking procedure, most vortices move long distances $u \gg a_{0}$, allowing a dynamic reorganization.
On the other hand, vortices that enter and exit the sample crossing its edge during the shaking procedure are present only in a thin micrometric annulus near the sample boundary (less than 2\% of the sample area) and may only marginally affect the average bulk structure.

The claim that the small ac field $h_{ac}=2.5$~mOe, applied during the linear ac susceptibility measurements, does not modify the VL configuration, i.e. the technique is non-invasive, has been extensively discussed in previous publications \cite{Pasquini2008}: the linear non-dissipative Campbell regime \cite{Campbell} holds when each vortex performs small harmonic oscillations around a minimum of  the effective pinning potential well. These small oscillations propagate inside the sample up to the Campbell penetration depth, generally smaller
than the sample dimension. The same reasons outlined above (replacing $h_{sh}\sim 7$~Oe by $h_{ac}=2.5$~mOe) lead to the conclusion that during the ac measurement even outmost vortices move typically less than $3$ nm $<\xi_{0}<a_{0}$, where $\xi_0$ is the coherence length, not modifying the VL configuration.

\subsection{Linear and non-linear ac regimes: ac field penetration and dissipation}

\begin{figure}[ht!]
\centering
\includegraphics[width=\linewidth]{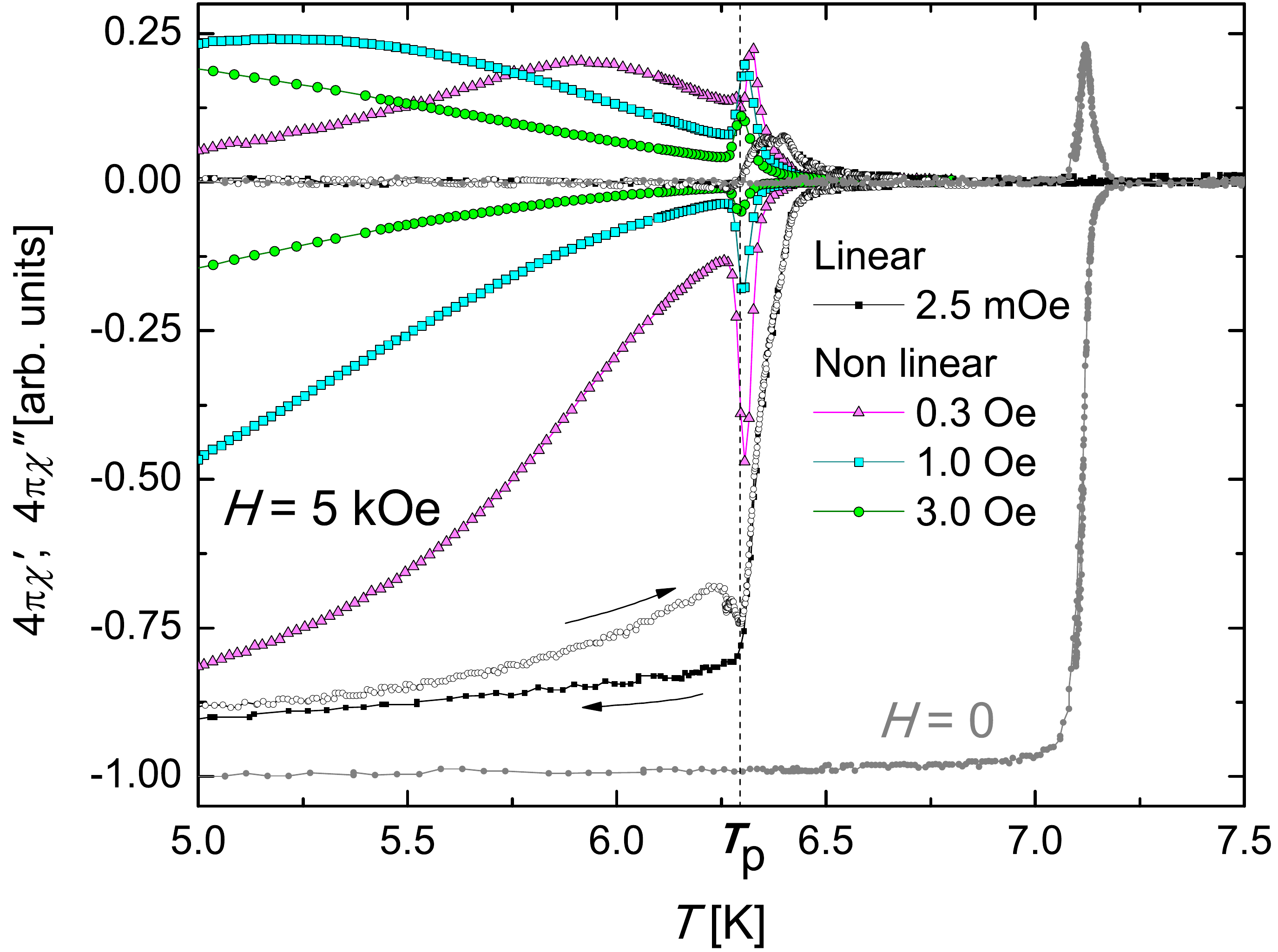}
\caption{Linear (black symbols) and non-linear (color symbols) ac response for different ac amplitudes, measured at $H=5$ kOe. The sharp zero-field transition is also included in light grey symbols. In the linear response ($h_{ac}=2.5$~mOe) $\chi''\sim 0$ below $T_{p}$, consistently with a non-invasive Campbell regime. The highest-amplitude ac field fully penetrates the sample, for $T>5$~K. Arrows in the linear curves show the direction of temperature variations. Non linear curves were recorded in cooling procedures.}
\label{fig:s2}
\end{figure}

The full penetration of the shaking field was previously confirmed by measuring the non-linear ac response in sample C, and results are shown is Figure \ref{fig:s2} (color curves): Below $T_{p}$, the normalized ac susceptibility is amplitude dependent, with a maximum out of phase component $\chi_{\max}''\sim 0.24$, characteristic of the ac Bean Critical response \cite{ClemySanchez}. At $T>5$~K, an ac field with amplitude $h_{ac}=3$~Oe (the maximum available in the MPMS) and frequency $f=1$~kHz fully penetrates the sample ($\chi' \sim 0$);  an even higher penetration is expected for the shaking field $h_{sh}\sim 7$~Oe. 
Notice that, despite the smaller volume of sample C, because samples A and C have exactly the same thickness $d$, similar non linear ac field penetration will hold \cite{ClemySanchez}. In the same figure, the linear (i.e. amplitude independent) ac susceptibility, measured in sample A with the SANS setup, is also included. Below $T_{p}$, the dissipative out-of-phase component $\chi''\approx 0$, as expected in the linear Campbell regime \cite{Pasquini2008,Campbell}.

\subsection{Magnetic field penetration in dc measurements}

\begin{figure}[ht!]
\centering
\includegraphics[width=\linewidth]{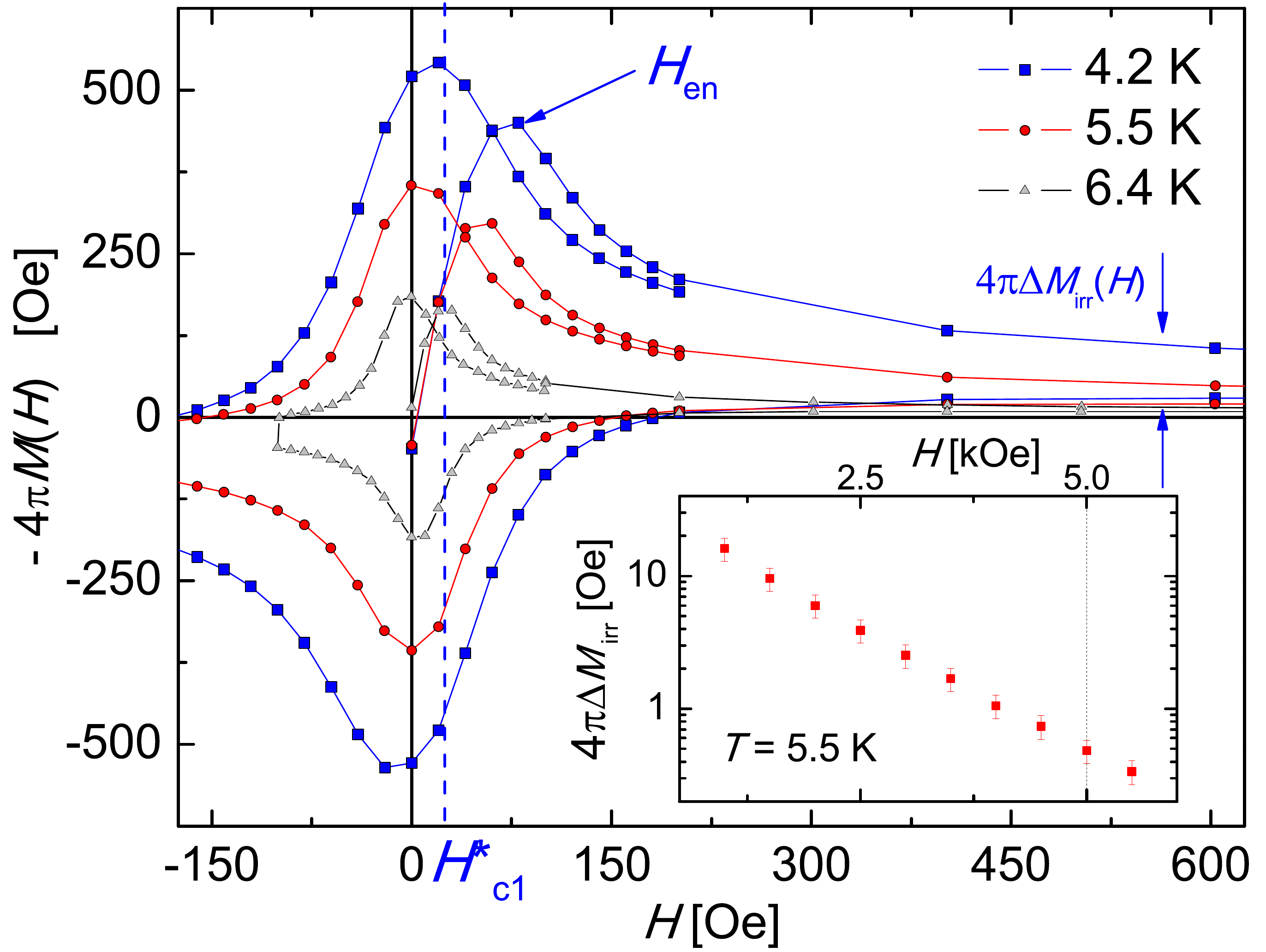}
\caption{(a) Magnetization loops $M(H)$ at low fields, for several temperatures. The shape is characteristic of low pinning thin superconductors in a transverse geometry; geometrical barriers inhibit the magnetic field entry up to $H_{en} > H_{c1}^{\ast }$ (the effective lower critical field). Inset: At $H \gg H_{en}$, a small irreversible magnetization $M_{irr}(H)$, arising from weak bulk pinning, persists. At $H=5$ k Oe, $4\pi M_{irr}<1$~Oe, consistently with the full bulk penetration of $h_{sh}\sim 7$~Oe.}
\label{fig:s3}
\end{figure}

The conclusions of the above paragraph are also consistent with results obtained by magnetization measurements. Figure \ref{fig:s3} shows examples of magnetization curves at low fields $H$ measured at different temperatures. The shape of the $M(H)$ loops is characteristic of low pinning thin superconductors in a transverse geometry \cite{Brandt99}, where geometrical barriers inhibit the magnetic field entrance up to a field $H_{en}$ higher than the effective lower critical field $H_{c1}^{\ast }$ ($\sim 0.15\ H_{c1}$ in our experimental geometry). At fields $H \gg H_{en}$, geometrical barriers do not play any significant role but a small irreversible magnetization $M_{irr}(H)$, arising from weak bulk pinning, persists. It can be observed in the inset of figure \ref{fig:s3} that $M_{irr}$ decays quickly with increasing field. At $H=5$~kOe (the magnetic field applied in our SANS experiments), the low irreversibility ($4\pi M_{irr} < 1$~Oe) is consistent with the full bulk penetration of the shaking field $h_{sh}\sim 7$~Oe.

\begin{thebibliography}{99}

\bibitem{packing} C. P. Goodrich, A. J. Liu, and S. R. Nagel, Nat. Phys. \textbf{10}, 578 (2014).

\bibitem{Suderow2014} I. Guillamón, R. Córdoba, J. Ses{\'e}, J. M. De Teresa, M. R. Ibarra, S. Vieria, and H. Suderow, Nat. Phys. \textbf{10}, 851 (2014).

\bibitem{BG1} T. Giamarchi and P. Le Doussal, Phys. Rev. B \textbf{52}, 1242 (1995).

\bibitem{BG2} T. Klein, I. Joumard, S. Blanchard, J. Marcus, R. Cubitt, T Giamarchi, and P. Le Doussal, Nature (London) \textbf{413}, 404 (2001).

\bibitem{Andrei2007} X. Du, G. Li, E. Y. Andrei, M. Greenblatt, and P. Shuk, Nat. Phys. \textbf{3}, 111 (2007).

\bibitem{ODT1} T. Giamarchi and P. Le Doussal, Phys. Rev. B \textbf{55}, 6577 (1997).

\bibitem{ODT2} A. M. Troyanovski, M. van Hecke, N. Saha, J. Aarts, and P. H. Kes, Phys. Rev. Lett. \textbf{89}, 147006 (2002).

\bibitem{ODTN1} P. L. Gammel, U. Yaron, A. P. Ramirez, D. J. Bishop, A. M. Chang, R. Ruel, L. N. Pfeiffer, E. Bucher, G. D'Anna, D. A. Huse, K. Mortensen, M. R. Eskildsen, and P. H. Kes, Phys. Rev. Lett. \textbf{80}, 833 (1998).

\bibitem{HE1} W. Henderson, E. Y. Andrei, M. J. Higgins, and S. Bhattacharya, Phys. Rev. Lett. \textbf{77}, 2077 (1996).

\bibitem{Paltiel2000} Y. Paltiel, E. Zeldov, Y. N. Myasoedov, H. Shtrikman, S. Bhattacharya, M. J. Higgins, Z. L. Xiao, E. Y. Andrei, P. L. Gammel, and D. J. Bishop, Nature (London) \textbf{403}, 398 (2000).

\bibitem{Pasquini2008} G. Pasquini, D. P{\'e}rez Daroca, C. Chiliotte, G. S. Lozano, and V. Bekeris, Phys. Rev. Lett. \textbf{100}, 247003 (2008).

\bibitem{Daroca2011} D. P{\'e}rez Daroca, G. Pasquini, G. S. Lozano, and V. Bekeris, Phys. Rev. B \textbf{84}, 012508 (2011).

\bibitem{SN} H. A. Hanson, X. Wang, I. K. Dimitrov, J. Shi, X. S. Ling, B. B. Maranville, C. F. Majkrzak, M. Laver, U. Keiderling, and M. Russina, Phys. Rev. B \textbf{84}, 014506 (2011).

\bibitem{SN2} A. Pautrat, M. Aburas, Ch. Simon, P. Mathieu, A. Br\^ulet, C. D. Dewhurst, S. Bhattacharya, and M. J. Higgins, Phys. Rev. B \textbf{79}, 184511 (2009).

\bibitem{Menon2012} G. I. Menon, Phys. Rev. B \textbf{65}, 104527 (2002); G. I. Menon, G. Ravikumar, M. J. Higgins, and S. Bhattacharya, Phys. Rev. B \textbf{85}, 064515 (2012).

\bibitem{Xiao2000} Z. L. Xiao E. Y. Andrei, P. Shuk, and M. Greenblatt, Phys. Rev. Lett. \textbf{85}, 3265 (2000).

\bibitem{Xiao2004} Z. L. Xiao, O. Dogru, E. Y. Andrei, P. Shuk, and M. Greenblatt, Phys. Rev. Lett. \textbf{92}, 227004 (2004).

\bibitem{Forgan2008} M. Laver, E. M. Forgan, A. B. Abrahamsen, C. Bowell, Th. Geue, and R. Cubitt, Phys. Rev. Lett. \textbf{100}, 107001 (2008).

\bibitem{Giamarchi1995} T. Giamarchi and P. Le Doussal, Phys. Rev. Lett. \textbf{75}, 3372 (1995).

\bibitem{ReviewMorten} M. R. Eskildsen, Front. Phys. \textbf{6}, 398 (2011).

\bibitem{BellLabs} C.S. Oglesby, E. Bucher, C. Kloc, and H. Hohl, J. Cryst. Growth \textbf{137}, 289 (1994).

\bibitem{Colo2014} B. Raes, C. C. de Souza Silva, A. V. Silhanek, L. R. E. Cabral, V. V. Moshchalkov, and J. Van de Vondel, Phys. Rev. B \textbf{90}, 134508 (2014).

\bibitem{Campbell} A. M. Campbell, J. Phys. C \textbf{4}, 3186 (1971).

\bibitem{Yaron1995} U. Yaron, P. L. Gammel, D. A. Huse, R. N. Kleiman, C. S. Oglesby, E. Bucher, B. Batlogg, D. J. Bishop, K. Mortensen, and K. N. Clausen, Nature (London) \textbf{376}, 753 (1995).

\bibitem{lambdaT} $\lambda (T)$ has been estimated from: J. D. Fletcher, A. Carrington, P. Diener, P. Rodière, J. P. Brison, R. Prozorov, T. Olheiser, and R. W. Giannetta, Phys. Rev. Lett. \textbf{98}, 057003 (2007).

\bibitem{Brandt1995} E. H. Brandt, Rep. Prog. Phys. \textbf{58} 1465 (1995).

\bibitem{Ganguli2014}  S. C. Ganguli, H. Singh, G. Saraswat, R. Ganguly, V. Bagwe, P. Shirage, A. Thamizhavel, and P. Raychaudhuri,  Sci. Rep. \textbf{5}, 10613 (2015).

\bibitem{Marchevsky2001} M. Marchevsky, M. J. Higgins, and S. Bhattacharya, Nature (London) \textbf{409}, 591 (2001).

\bibitem{Daniilidis2007} N. D. Daniilidis, S. R. Park, I. K. Dimitrov, J. W. Lynn, and X. S. Ling, Phys. Rev. Lett. \textbf{99}, 147007 (2007).

\bibitem{Pine2008} L. Corté, P. M. Chaikin, J. P. Gollub, and D. J. Pine, Nat. Phys. \textbf{4}, 420 (2008).

\bibitem{Reichhardt2009} C. Reichhardt and C. J. Olson Reichhardt, Phys. Rev. Lett. \textbf{103}, 168301 (2009).

\bibitem{Mullock} S. J. Mullock and J. E. Evetts, J. Appl. Phys. \textbf{57},2588 (1985).

\bibitem{TheoWorks} M. C. Marchetti and D. Nelson, Phys. Rev. B \textbf{41},1910 (1990); J. Kierfeld and V. Vinokur, Phys. Rev. B \textbf{61}, R14928 (2000).

\bibitem{Kolton2000} A. B. Kolton, D. Domínguez, C. J. Olson, and N. Gr\o nbech-Jensen, Phys. Rev. B \textbf{62}, R14657 (2000).

\bibitem{Yaron1994} U. Yaron, P. L. Gammel, D. A. Huse, R. N. Kleiman, C. S. Oglesby, E. Bucher, B. Batlogg, D. J. Bishop, K. Mortensen, K. Clausen, C. A. Bolle, and F. De La Cruz,  Phys. Rev. Lett. \textbf{73}, 2748 (1994).

\bibitem{grasp} C. Dewhurst, computer code \textsc{grasp}: Graphical Reduction and Analysis SANS Program, version 6.93b, 2014.

\bibitem{octave} John W. Eaton, David Bateman, and Søren Hauberg, \textit{GNU Octave version 3.0.1 manual: a high-level interactive language for numerical computations} (CreateSpace, 2009).

\bibitem{morten} M. R. Eskildsen,  Ph.D. thesis, Ris\o National Laboratory and University of Copenhagen, Denmark, 1998.

\bibitem{ClemySanchez} J. R. Clem and A. Sanchez, Phys. Rev. B \textbf{50}, 9355 (1994).

\bibitem{Brandt99} E. H. Brandt, Phys Rev B \textbf{60}, 11939 (1999).

\end{thebibliography}
\end{document}